\documentclass[12pt]{article}
\usepackage{eqsection,latexsym,epsf,cite,amssymb,multirow,wasysym,graphicx}
\usepackage[english]{babel}

\begin{document} 

\setlength{\unitlength}{0.2cm}

\title{Geometrical Properties of 
Two-Dimensional Interacting Self-Avoiding Walks at the
$\theta$-Point
}
\author{
  \\
  {\small Sergio Caracciolo} \\[-0.2cm]
  {\small\it Dipartimento di Fisica and INFN --- Sezione di Milano I} \\[-0.2cm]
  {\small\it Universit\`a degli Studi di Milano} \\[-0.2cm]
  {\small\it Via Celoria 16, I-20133 Milano, Italy} \\[-0.2cm]
  {\small e-mail: {\tt Sergio.Caracciolo@mi.infn.it}} \\[-0.2cm]
        \\[-0.1cm]  \and
  {\small Marco Gherardi} \\[-0.2cm]
  {\small\it Dipartimento di Fisica and INFN --- Sezione di Milano I} \\[-0.2cm]
  {\small\it Universit\`a degli Studi di Milano} \\[-0.2cm]
  {\small\it Via Celoria 16, I-20133 Milano, Italy} \\[-0.2cm]
  {\small e-mail: {\tt Marco.Gherardi@mi.infn.it}} \\[-0.2cm]
        \\[-0.1cm]  \and
  {\small Mauro Papinutto} \\[-0.2cm]
  {\small\it Laboratoire de Physique Subatomique et de Cosmologie,} \\[-0.2cm] 
  {\small\it UJF/CNRS-IN2P3/INPG, 53 rue des Martyrs, F-38026 Grenoble, 
     France} \\[-0.2cm]
  {\small e-mail: {\tt Mauro.Papinutto@lpsc.in2p3.fr}}     \\[-0.2cm]
  \\[-0.1cm]  \and
  {\small Andrea Pelissetto} \\[-0.2cm]
  {\small\it Dipartimento di Fisica and INFN -- Sezione di Roma I} \\[-0.2cm]
  {\small\it Universit\`a degli Studi di Roma ``La Sapienza"} \\[-0.2cm]
  {\small\it P.le A.Moro 2, I-00185 Roma, Italy} \\[-0.2cm]
  {\small e-mail: {\tt Andrea.Pelissetto@roma1.infn.it}}   \\[-0.2cm]
  {\protect\makebox[5in]{\quad}}  
  \\
}

\maketitle
\thispagestyle{empty}   


\begin{abstract}
We perform a Monte Carlo simulation of two-dimensional $N$-step
interacting self-avoiding walks at the $\theta$ point, with lengths
up to $N=3200$. We compute the critical exponents, verifying the 
Coulomb-gas predictions, the $\theta$-point temperature 
$T_\theta = 1.4986(11)$, and several invariant size ratios.
Then, we focus on the geometrical features of the walks,
computing the instantaneous shape ratios, the average asphericity,
and the end-to-end distribution function. For the latter quantity, 
we verify in detail the theoretical predictions for its small- and 
large-distance behavior.
\end{abstract}

\clearpage

\newcommand{\be}{\begin{equation}}
\newcommand{\ee}{\end{equation}}
\newcommand{\beq}{\begin{equation}}
\newcommand{\eeq}{\end{equation}}
\newcommand{\bea}{\begin{eqnarray}}
\newcommand{\eea}{\end{eqnarray}}
\newcommand{\<}{\langle}
\renewcommand{\>}{\rangle}

\def\spose#1{\hbox to 0pt{#1\hss}}
\def\ltapprox{\mathrel{\spose{\lower 3pt\hbox{$\mathchar"218$}}
 \raise 2.0pt\hbox{$\mathchar"13C$}}}
\def\gtapprox{\mathrel{\spose{\lower 3pt\hbox{$\mathchar"218$}}
 \raise 2.0pt\hbox{$\mathchar"13E$}}}
\def\lle{<<}
\def\gge{>>}

\newcommand{\R}{\hbox{{\rm I}\kern-.2em\hbox{\rm R}}}
\def\brho{\mbox{\protect\boldmath $\rho$}}

\newcommand{\reff}[1]{(\ref{#1})}
\def\smfrac#1#2{{\textstyle\frac{#1}{#2}}}

\section{Introduction}

Self-avoiding walks (SAWs) have been extensively studied during the years.
They have a rich mathematical structure and represent one of the simplest 
models of critical behavior.
Moreover, they are also relevant for the understanding of 
the universal features of large-molecular-weight macromolecules in solution
\cite{deGennes-72,Daoud-75,DC-75,Emery-75,%
deGennes-79,Freed-87,dCJ-book,Schaefer-99}.
Indeed, the study of the asymptotic behavior of long SAWs (and also of walks 
with different architecture, like self-avoiding rings or stars) 
allows one to obtain 
predictions for several structural and thermodynamical 
properties --- for instance critical exponents, structure factors, 
osmotic pressure, etc. --- of polymers both in dilute and 
in semidilute good-solvent 
solutions in the limit of infinite degree of polymerization.
These predictions are in good agreement with the (usually much less 
precise) experimental results.
By adding an attractive interaction SAWs also allow us to study the 
critical transition ($\theta$-point) between good-solvent and poor-solvent
behavior \cite{deGennes-79,Freed-87,dCJ-book,Schaefer-99}. 
The phase diagram of interacting SAWs is well-known. 
Above the $\theta$-temperature 
$T_\theta$, the large-$N$ behavior is 
temperature-independent and in the same 
universality class as that observed for athermal SAWs, 
which correspond to the case $T=\infty$. 
On the other hand, for $T < T_\theta$ typical walks are compact:
this is the so-called collapsed or poor-solvent regime. 
At the $\theta$ temperature, interacting SAWs show a quite interesting 
tricritical behavior. The two-dimensional case has been extensively studied
by Coulomb-gas techniques and conformal field theory (CFT). They have 
provided exact predictions for the critical exponents \cite{DS-87},
which have been confirmed by several high-precision numerical studies
\cite{Ishinabe_87,SS-88,MV-90,ML-90,FOT-92,CM-93,OPBG-94,GH-95,Nidras_96,
BWEGGLOW-98,NKMR-01,GV-09,LKL-10}. 
Also the collapsed phase has been discussed in detail
\cite{OPB-93,Owczarek-93,GH-02,HG-02,BOS-06}, including the crossover
behavior close to the $\theta$ point \cite{OP-03}.

In this paper we perform a Monte Carlo (MC) simulation of two-dimensional
interacting SAWs close to the $\theta$ point, with the purpose of 
determining some geometrical features of the walks. In particular,
we shall focus on the instantaneous shape of the walks and on
the end-to-end distribution function (EEDF).
As is well known, walks are not instantaneously spherical. Their
shape is usually characterized by considering combinations of 
the eigenvalues of the 
gyration tensor. One of them, the mean asphericity, is
 an essential ingredient in theoretical studies of the 
hydrodynamic behavior of dilute polymer solutions 
\cite{Kuhn-34,Kramers-46,ABCK-80}.
The EEDF has been
extensively studied in three-dimensions, because of its 
theoretical interest \cite{FH-61,Fisher-66,Mazur-65,MM-71,McKenzie-76,%
desCloizeaux-74_80,Bishop-Clarke_91,Bishop-etal_91,Dayantis-Palierne_91,%
Eizenberg-Klafter_93,Pedersen-etal_96,Valleau_96,CCP-00}. 
Under the mapping of SAWs onto the $n\to 0$ 
$\sigma$ model (or $\lambda \phi^4$ theory), it corresponds to the 
spin-spin correlation function, which is the basic object 
of field-theoretical calculations. Moreover, it can be also
accessed experimentally, by performing neutron-scattering
experiments  on solutions of end-marked polymers \cite{dCJ-book}. 

In this paper we will obtain a high-precision determination of the 
EEDF, which will be compared with several predictions obtained
by using the standard mapping onto the $n=0$ spin model and with 
phenomenological expressions that have been shown to be quite accurate 
in three dimensions. As for the shape of the walks, we will determine the 
average asphericity, comparing it with results obtained in the 
good-solvent regime \cite{BCRF-91,vFYMB-09} 
and for noninteracting random walks \cite{DE-89}.

The paper is organized as follows. In Sec.~\ref{sec2} 
we define the interacting SAWs and some basic quantities.
In Sec.~\ref{sec3} we use our MC results to determine the 
$\theta$ temperature and verify the theoretical predictions 
for the critical exponents. In Sec.~\ref{sec4} we study
the walk shape and in Sec.~\ref{sec5} the EEDF.
Finally, in Sec.~\ref{sec6} we summarize our conclusions.

\section{Model and definitions} \label{sec2}

In this paper we consider self-avoiding walks (SAWs) on a 
two-dimensional square lattice. An $N$-step  SAW $\omega$ is a set
of $N+1$ lattice sites $\omega_0 = 0$, $\omega_1$, $\ldots$,
$\omega_N$, such that $\omega_i$ and $\omega_{i+1}$ are lattice
nearest neighbors. For each walk, we define the energy $\cal E$
as follows:
\be
{\cal E} \equiv - \sum_{i=0}^{N-3} \sum_{j=i+3}^N c_{ij},
\ee 
where
\be
c_{ij} \equiv \cases{1 & $\qquad$ if $|\omega_i - \omega_j| = 1$; \cr
                0 & $\qquad$ otherwise.
               }
\ee
Essentially, $\cal E$ is the number of nearest-neighbor contacts 
without considering the trivial contacts between subsequent monomers.

We consider the ensemble of $N$-step walks with partition function
\be
Z_N = \sum_{\{\omega\}} e^{-\beta {\cal E}},
\ee
where $\beta$ is the inverse temperature and the sum is extended over 
all $N$-step walks. We will study the behavior close to the 
$\theta$ temperature $\beta_\theta$.
For the model we consider, the best present-day estimates of $\beta_\theta$ 
on the square lattice 
are reported in Table \ref{beta-theta}.

\begin{table}
\begin{center}
\begin{tabular}{|lccc|}
\hline\hline
& year & method & $\beta_\theta$ \\
\hline
Ref. \protect\cite{Ishinabe_87} & 1987 & EE & 0.75 \\
Ref. \protect\cite{SS-88}       & 1988 & MC & 0.65(3) \\
Ref. \protect\cite{MV-90}       & 1990 & EE & 0.67(4) \\
Ref. \protect\cite{ML-90}       & 1990 & MC & 0.658(4) \\
Ref. \protect\cite{FOT-92}      & 1992 & EE & 0.657(16) \\
Ref. \protect\cite{CM-93}       & 1993 & MC & 0.658(4) \\
Ref. \protect\cite{OPBG-94}     & 1994 & EE & 0.660(5) \\
Ref. \protect\cite{GH-95}       & 1995 & MC & 0.665(2) \\
Ref. \protect\cite{Nidras_96}   & 1996 & MC & 0.664, 0.666 \\
Ref. \protect\cite{Bastolla_98} & 1997 & MC & 0.667(1) \\
Ref. \protect\cite{CNDC-09}     & 2009 & MC & 0.664(8) \\
this work                       & 2010 & MC & 0.6673(5) \\
\hline\hline
\end{tabular}
\end{center}
\caption{Estimates of $\beta_\theta$ on the square lattice.
EE stands for exact enumeration, MC for Monte Carlo.
}
\label{beta-theta}
\end{table}

We consider three different observables that measure the size of the walk:
\begin{itemize}
\item the square end-to-end distance
\be
\label{raggioendtoend}
R_e^2 \equiv (\omega_N - \omega_0)^2\; ;
\ee
\item the square radius of gyration
\be
\label{raggiogirazione}
R_g^2 \equiv \frac{1}{N+1}\sum_{i=0}^{N}\Bigg(\omega_i -
\frac{1}{N+1}\sum_{k=0}^N \omega_k\Bigg)^2 =
\frac{1}{2(N+1)^2}\sum_{i,j=0}^{N}(\omega_i - \omega_j)^2 \; ;
\ee
\item the square monomer distance from an endpoint
\be
\label{raggioquadra}
R_m^2 \equiv \frac{1}{N+1}\sum_{i=0}^{N}(\omega_i-\omega_0)^2 \; .
\ee
\end{itemize} 
Correspondingly, we define the universal ratios
\be
A_N \equiv {\<R_g^2\>_N\over \<R_e^2\>_N}, \qquad
B_N \equiv {\<R_m^2\>_N\over \<R_e^2\>_N}, \qquad
C_N \equiv {\<R_g^2\>_N\over \<R_m^2\>_N}, 
\ee
and the combination (the exponents $\gamma_\theta$ and 
$\nu_\theta$ are defined below)
\be
F_N \equiv \left(2 + {2\over \gamma_\theta + 2 \nu_\theta}\right) A_N 
      - 2 B_N + {1\over2} = {23\over 8} A_N - 2 B_N + {1\over2}.
\label{def-F}
\ee
For two-dimensional non-interacting SAWs it has been proved
\cite{CS_1989_JPhysA, CPS_1990_JPhysA}
that the corresponding $F_N$ (with the appropriate 
exponents $\gamma$ and $\nu$) 
vanishes in the limit $N\to\infty$.
It has been conjectured and verified numerically
\cite{OPBG-94} that $F_\infty = 0$ also at the $\theta$ point.

In a neighborhood of $\beta_\theta$ the radii have a scaling behavior of the 
form
\be
\< R^2\>_N = N^{2\nu_\theta} f(x) \qquad 
x \equiv N^\phi (\beta - \beta_\theta),
\label{scalingR2}
\ee
with $f(0)\not=0$. More precisely, this scaling form is valid 
for $\beta\to\beta_\theta$, $N\to\infty$ at fixed $x$.
In two dimensions, CFT and Coulomb-gas techniques allow us to 
compute the universal exponents 
$\phi$ and $\nu_\theta$. They are given by \cite{DS-87}
\be
\nu_\theta = {4\over7}, \qquad \phi = {3\over7}.
\ee
The crossover exponent $\phi$ can be measured directly at $\beta = \beta_\theta$
by considering the specific heat $h_N$, which scales as 
\be
h_N \equiv {1\over N}\left(\< {\cal E}^2 \>_N - \< {\cal E} \>^2_N \right)
    \sim N^{2\phi -1},
\ee
or the temperature dependence of the radii, 
\be
D_N \equiv - {1\over \< R^2\>_N} {d \< R^2\>_N\over d\beta}  
  = \< {\cal E} \>_N - {\< R^2 {\cal E} \>_N \over \< R^2\>_N} 
  \sim N^\phi.
\ee
In the analysis of the end-to-end distribution function we will also 
consider the exponent $\gamma_\theta$, which controls the large-$N$
behavior of the partition function at the $\theta$ temperature,
\be
Z_N \sim \mu^N N^{\gamma_\theta - 1},
\ee
where $\mu$ is a lattice- and model-dependent constant. 
The exponent $\gamma_\theta$ is universal; CFT and Coulomb-gas 
calculations predict \cite{DS-87}
$\gamma_\theta = 8/7$. 

\begin{table}
\begin{center}
\begin{tabular}{|cccccc|}
\hline\hline
$N$ & $\<R^2_g\>_N$ & $\<R^2_e\>_N$ & $\<R^2_m\>_N$ &
$\<{\cal E}\>_N$ & $h_N$ \\
\hline
100 & 43.438(8) & 241.87(7) & 123.02(3) &
46.304(2) & 68.10(2) \\
800 & 472.72(30) & 2604(3) & 1332(1) & 429.712(26) & 909.53(60) \\    
1600 & 1050.19(37) & 5784(3) & 2957(1) & 877.717(21) & 2034.49(68) \\
3200 & 2338.7(1.5) & 12895(13) & 6584(5) & 1780.259(53) & 4460.2(2.5) \\
\hline
\hline
$N$ & $A_N$ & 
$B_N$ & $C_N$ & $D_N$ & $F_N$\\
\hline
100 & 0.179592(61) & 0.50863(18) & 0.35309(97) & 2.2793(15) & $-$0.0009(4) \\
800 & 0.18152(22) & 0.51139(64) & 0.35496(35) & 6.826(15) &   $-$0.0009(14)\\
1600 & 0.18156(12) & 0.51123(35) & 0.35514(19) & 9.54(1) &    $-$0.0005(8)\\
3200 & 0.18137(21) & 0.51060(63) & 0.35521(35) & 13.170(29)&  
   \hphantom{$-$}0.0002(14)\\
\hline\hline
\end{tabular}
\caption{Estimates of $\<R^2_g\>_N$, $\<R^2_e\>_N$, $\<R^2_m\>_N$,
$\<{\cal E}\>_N$, $h_N$,
$A_N$, $B_N$, $C_N$, $D_N$, and $F_N$. }
\label{table-risultati}
\end{center}
\end{table}

\section{Determination of the critical exponents} \label{sec3}

We performed a MC simulation using the extended reptation algorithm
discussed in detail in Refs.~\cite{Caracciolo-etal_2000,CPP-02}.
We fixed $\beta = 0.665$, which is the 
estimate of $\beta_\theta$ presented in Ref.~\cite{GH-95}, 
and performed runs for 
$N=100,800,1600,3200$. Some results are reported in 
Table \ref{table-risultati}. Since $\beta_\theta$ is not exactly known,
we also computed several quantities
for $\beta = 0.665 - 0.0005n$, $n=1,2,3,4$, using the standard reweighting
method.

\begin{table}
\begin{center}
\begin{tabular}{|cccccc|}
\hline\hline
& $N_{\rm min}$ & $\nu$ & $\phi$ & $\beta_\theta$ & $\beta_\theta$ (f.exp.)  \\
\hline
\multirow{2}{*}{$R_g$}
& 100 & 0.5720(2) & 0.480(4) & 0.6669(1) & 0.6675(1) \\
& 800 & 0.5721(2) & 0.480(5) & 0.6669(1) & 0.6675(1) \\
\hline
\multirow{2}{*}{$R_e$}
& 100 & 0.5678(3) & 0.480(5) & 0.6677(1) & 0.6670(1) \\
& 800 & 0.5683(3) & 0.480(5) & 0.6676(1) & 0.6670(1) \\
\hline
\multirow{2}{*}{$R_m$}
& 100 & 0.5706(3) & 0.478(6) & 0.6672(1) & 0.6671(1) \\
& 800 & 0.5711(3) & 0.478(5) & 0.6670(1) & 0.6671(1) \\
\hline\hline
\end{tabular}
\caption{Estimates of the exponents and of $\beta_\theta$ obtained 
by fitting the reweighted data to Eq.~(\ref{fit-radii}). 
The critical value $\beta_\theta$ in the rightmost column
[$\beta_\theta$ (f.exp.)] is obtained by fixing the theoretical values 
for the exponents, i.e., by fitting the data to 
 $\<R^2\>_N = a N^{8/7} + b N^{11/7} (\beta - \beta_\theta)$.
}
\label{table-reweighting}
\end{center}
\end{table}

In order to determine $\nu_\theta$ we perform fits of the radii.
If we fit all data at $\beta_\theta = 0.665$ to $a N^{2\nu_\theta}$ 
we obtain $\nu_\theta\approx 0.573$, while, if we discard the results
corresponding to $N = 100$, we obtain $\nu_\theta\approx 0.577$.
These estimates are close to the theoretical 
prediction $\nu_\theta = 4/7 \approx 0.5714\ldots$. The slight discrepancy
is probably due to the fact that $\beta = 0.665$ is slightly smaller than the 
$\theta$-value $\beta_\theta$, so that we are seeing the 
beginning of the crossover towards the good-solvent value $\nu = 3/4$.
A better analysis consists in fitting the data at $\beta = 0.665$ and the 
reweighted data to Eq.~(\ref{scalingR2}). Since $x$ is small for our data,
assuming that the function $f(x)$ is regular at $x = 0$, we can expand it
in powers of $x$. At first order we obtain the scaling form
\be
\<R^2\>_N = a N^{2\nu_\theta} + b N^{2\nu_\theta + \phi} (\beta - \beta_\theta),
\label{fit-radii}
\ee
valid for $x\equiv N^\phi (\beta - \beta_\theta)\ll 1$.
We first perform fits taking $a$, $b$, $\nu_\theta$, $\phi$, and 
$\beta_\theta$ as free parameters. The results are 
reported in Table \ref{table-reweighting}, as a function of $N_{\rm min}$,
the minimum length allowed in the fits. The results obtained in the 
analysis of the three ratios are reasonably close and indicate 
\begin{eqnarray}
\nu_\theta  &=&   0.570(2), \\
\phi        &=&   0.479(6),   \label{phi-est}\\
\beta_\theta&=&   0.6673(5),  \label{betatheta-est} \\
T_\theta    &=&  1/\beta_\theta = 1.4986(11).
\end{eqnarray}
These estimates are the average of all fits and the reported error 
is such to include all results and the corresponding errors.
The exponent $\nu_\theta$ is in good agreement with the 
Coulomb-gas prediction
$\nu_\theta = 4/7\approx 0.571$.
On the other hand, the exponent $\phi$ is significantly larger than 
$\phi = 3/7 \approx 0.429$. This may be due to neglected scaling corrections
and/or to the neglected terms in the expansion of the function $f(x)$,
although, we must admit, we have no evidence of corrections in our 
results, which are stable with respect to $N_{\rm min}$ and 
to the observable considered.

The estimate of $\beta_\theta$ is in good agreement with the most recent one 
obtained in Ref.~\cite{Bastolla_98}, 
in which much longer walks were used. 
The fit in which we assume the theoretical values of the exponents gives 
estimates of $\beta_\theta$ that have a significantly smaller statistical 
error, see Table \ref{table-reweighting}. 
However, note that we neglect here scaling corrections
which can --- and probably do, 
given the observed discrepancy for the exponent $\phi$ ---
give rise to systematic deviations which are larger than the tiny 
statistical errors.
Hence, we shall keep the conservative estimate (\ref{betatheta-est}).

In order to obtain independent estimates of the crossover exponent, 
we also analyze $D_N$ and $h_N$. By fitting the MC data we obtain
$\phi = 0.450(4)$ (from $h_N$) and $\phi = 0.436(5)$ (from $D_N$).
They differ significantly from the estimate (\ref{phi-est}),
and thus provide evidence that the apparent stability 
observed in the fits of the radii and, therefore, the relatively small
error, should not be trusted.
The estimate from the analysis of $h_N$ and $D_N$ are in better agreement 
with the theoretical value than the estimate (\ref{phi-est}). 
The still present tiny discrepancies 
indicate that corrections to scaling and/or crossover effects are  relevant 
and give rise to systematic deviations, which are larger than the 
statistical errors. This is consistent with the results of Ref.~\cite{GH-95}
which found $\phi = 0.435(6)$ with significant scaling corrections.
Other recent estimates of $\phi$ are $\phi = 0.419(3)$ \cite{NKMR-01},
$\phi = 0.436(7)$ (in a model with explicit solvent) \cite{GV-09},
and $\phi = 0.422(12)$ from the analysis of the partition-function
zeroes \cite{LKL-10}.

Finally, we consider the invariant ratios 
$A_N$, $B_N$, $C_N$, and $F_N$. MC estimates at 
$\beta = 0.665$ are
reported in Table \ref{table-risultati}.
If we exclude that data with $N = 100$, they are constant within error
bars, indicating that scaling corrections and 
crossover effects are smaller than the statistical errors. 
Conservatively, we estimate the asymptotic value by averaging the results
with $N\ge 1600$. We obtain 
\begin{eqnarray}
A_\infty &=& 0.18151(10) \\
B_\infty &=& 0.51106(31) \\
C_\infty &=& 0.35516(17) \\
F_\infty &=& -0.0003(7).
\end{eqnarray}
The ratios $A_\infty$ and $B_\infty$ are in good agreement
with the results of Ref.~\cite{OPBG-94}:
$A_\infty = 0.180(1)$, $B_\infty = 0.510(2)$.
The estimate of $F_\infty$ is fully compatible with zero,
providing further support to the conjecture $F_\infty = 0$ 
of Ref.~\cite{OPBG-94}.

\section{Gyration tensor and asphericity} \label{sec4}

\begin{figure}
\begin{center}
\includegraphics{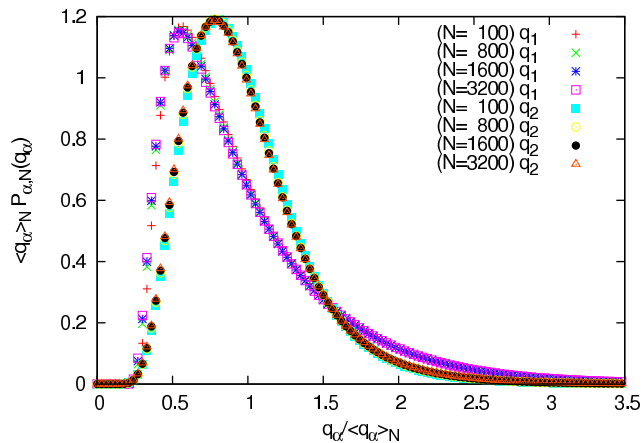}
\caption{Rescaled  distribution functions 
$\< q_\alpha\>_N P_{\alpha,N}(q_\alpha)$ for the two
eigenvalues $q_\alpha$ of the gyration tensor. }
\label{figure-eigenvalues}
\end{center}
\end{figure}

\begin{table}
\begin{center}
\begin{tabular}{|ccccc|}
\hline\hline
$N$ & $\<q_1\>_N$ & $\<q_2\>_N$ & $s_1$ & ${\cal A}$ \\
\hline
100 & 35.543(8)   & 7.8954(8)  & 0.8182(2) & 0.37668(8) \\
800 & 386.85(30)  & 85.871(34) & 0.8183(8) & 0.37569(28) \\
1600 & 859.93(37) & 190.262(41)& 0.8188(5) & 0.37653(16) \\
3200 & 1916.6(1.5)& 422.06(17) & 0.8195(8) & 0.37805(28) \\
\hline\hline
\end{tabular}
\caption{MC estimates of the eigenvalues of the gyration tensor, of 
the shape factor $s_1$, and of the mean asphericity.}
\label{table-gyration}
\end{center}
\end{table}

It has been known for many years that polymers are not instantaneously
spherical in shape \cite{Kuhn-34}. 
In order to characterize the shape
we consider the gyration tensor defined by
\be
Q_{N,\alpha\beta} \equiv 
  {1\over 2 (N+1)^2} \sum_{i,j=0}^N 
   (\omega_{i,\alpha} - \omega_{j,\alpha})
   (\omega_{i,\beta} - \omega_{j,\beta}),
\ee
which is such that ${\rm Tr}\, Q_N = R^2_{g}$. 
The tensor $Q_{N,\alpha\beta}$ is symmetric and positive definite, hence it has two
positive eigenvalues $q_1 \ge q_2$. For $N\to \infty$
they are expected to scale as 
\be
\< q_\alpha\>_N \approx B_\alpha N^{2\nu_\theta},
\label{eq-eigenvalues-scaling}
\ee
and to obey a scaling law of the form
\be
P_{\alpha,N} (q_\alpha) = {1\over \< q_\alpha\>_N} 
    F_\alpha\left({q_\alpha\over \< q_\alpha\>_N}\right).
\label{fscal-eigenvalues}
\ee
In Fig.~\ref{figure-eigenvalues} we report the rescaled eigenvalue
distributions. All points fall quite nicely onto two different 
universal curves, confirming the validity of the scaling form
(\ref{fscal-eigenvalues}). Note that, although the average 
eigenvalues differ approximately by a factor of 4.5, see 
Table~\ref{table-gyration}, the two distribution functions
are similar. They have a sharp peak for 
${q_\alpha/ \< q_\alpha\>_N} \approx
0.55$ ($\alpha = 1$) and $0.75$ ($\alpha = 2$), a long tail, and go to zero
sharply for ${q_\alpha/ \< q_\alpha\>_N} \approx 0.25$.

To characterize quantitatively the shape of $\theta$ walks 
we introduce the shape factors 
\be
s_1 \equiv {\< q_1\>_N\over \< R^2_g\>_N}, \qquad
s_2 \equiv {\< q_2\>_N\over \< R^2_g\>_N} = 1 - s_1, \qquad
r_{12} \equiv {\< q_1\>_N\over \< q_2\>_N} = {s_1\over 1 - s_1},
\ee 
and the mean asphericity
\be
{\cal A} \equiv {1\over 2} \sum_{\alpha} 
        \left\< {(q_\alpha - \bar{q})^2\over \bar{q}^2} \right\>_N
        = \left\< {(q_1 - q_2)^2\over (q_1 + q_2)^2} \right\>_N,
\ee
where $\bar{q} = \sum_\alpha q_\alpha/2 = R^2_{g}/2$.
For a disk we have $s_1 = 1/2$, ${\cal A} = 0$, while 
a rod  gives $s_1 = 1$, ${\cal A} = 1$.

At variance with the ratios $A_N$, $\ldots$, the quantities $s_1$ and ${\cal A}$
at $\beta = 0.665$ show a systematic drift with $N$, which may be an indication
of the crossover towards the asymptotic good-solvent value. To take it
into account, we use the expected scaling behavior close to the 
$\theta$ point:
\be
s_1, {\cal A} = f[(\beta-\beta_\theta) N^\phi].
\ee
Expanding the function $f(x)$ to first order we obtain 
\be
s_1, {\cal A} = a_1 + a_2 (\beta - \beta_\theta) N^\phi.
\label{fitratio}
\ee
This implies that we should fit our data to 
$a_1 + b N^\phi$. The parameter $a_1$ gives the $\theta$-point 
estimate of the universal ratio. Using $\phi = 3/7$, we obtain 
$s_1 = 0.8179(4)$ if we fit all data, and $s_1 = 0.8169(20)$ if we 
discard $N = 100$. Conservatively, we quote as final result
\be
s_1 = 0.817(2).
\ee
From $s_1$ we can compute the ratio of the average eigenvalues:
\be
   r_{12} = 4.46(6)
\ee
Note that the walks are elliptical, with the major axis being 
approximately a factor of two longer than the minor one.

The $N$ dependence of the average asphericity is not monotonic, 
and thus the fitting form (\ref{fitratio}) 
cannot describe the data up to $N=100$. 
We thus only fit the data satisfying $N\ge 800$, obtaining
\be
    {\cal A} = 0.3726(7). \label{asphericity}
\ee
Note that the error is purely statistical and thus it does not include 
the systematic uncertainty due to the scaling corrections. 
It is interesting to compare the results for the asphericity
with those obtained under good-solvent conditions \cite{BCRF-91,vFYMB-09}:
\be
  {\cal A}_{\rm GS} = 0.503(1).
\ee
Clearly, at the $\theta$-point walks are more symmetric 
than in the good-solvent regime. Note that our result is closer 
to the random-walk value \cite{DE-89}
\be
  {\cal A}_{\rm RW} = {5\over2} - {7\over 4}\zeta(3) \approx 0.3964.
\ee
Thus, also in two dimensions, $\theta$ point interacting SAWs 
can be reasonably described by random walks (in three dimensions 
interacting SAWs are effectively random walks in the limit $N\to \infty$,
although for finite $N$ there are quite strong logarithmic corrections),
as far as the shape is concerned.

\section{End-to-end distribution function} \label{sec5}

\subsection{Definitions}

If $c_N({\bf r})$ is the
number of SAWs starting at the origin and ending in ${\bf r}$,
we define the normalized end-to-end distribution function (EEDF) as
\be
P_N({\bf r}) = {c_N({\bf r})\over \sum_r c_N({\bf r})}\; .
\ee
The mean squared end-to-end distance is related to $P_N({\bf r})$ by
\be
    \< R^2_{e}\>_N = \sum_r |{\bf r}|^2 P_N({\bf r}).
\ee
In most of the studies of the EEDF one usually defines a 
correlation length $\xi$ which is trivially related to $R_{e}$:
\be
\xi^2 = {1\over 4} \<R_{e}^2\>_N.
\label{eq2.3}
\ee
In the following we shall always use $\xi$ to characterize the polymer size.
In the limit $N\to\infty$, $|{\bf r}|\to \infty$, 
with $|{\bf r}| N^{-\nu}$ fixed, the function $P_N({\bf r})$ has the
scaling form \cite{Fisher-66,MM-71,desCloizeaux-74_80}
\be
P_N({\bf r}) \approx {1\over \xi^2} f(\rho) 
   \left[1 + O(N^{-\Delta})\right],
\label{deffrho}
\ee
where $\brho = {\bf r}/\xi$, $\rho = |\brho|$, 
and $\Delta$ is a correction-to-scaling exponent.
By definition
\bea
   \int_0^\infty 2 \pi \rho d\rho\, f(\rho) &=& 1\; ,
\label{norm1} \\
   \int_0^\infty 2 \pi \rho^{3} d\rho\, f(\rho) &=& 4\; .
\label{norm2}
\eea
Several facts are known about $f(\rho)$. For large values of $\rho$ 
it behaves as 
\cite{FH-61,Fisher-66,MM-71,desCloizeaux-74_80}
\be
f(\rho) \,\approx f_\infty \rho^\sigma
   \exp\left(-D \rho^\delta\right)\; ,
\label{flargerho}
\ee
where $\sigma$ and $\delta$ are given by
\bea
   \delta &=& {1\over 1-\nu_\theta} = {7\over 3} \approx 2.33 , \label{delta} \\
   \sigma &=& {4 \nu_\theta - 2 \gamma_\theta \over 2 (1-\nu_\theta)} 
      = 0.
\eea
For $\rho\to 0$, we have \cite{MM-71,desCloizeaux-74_80} instead
\be
   f(\rho) \approx f_0 \rho^\theta,
\label{fsmallrho}
\ee
where
\be
\theta =\, {\gamma_\theta - 1\over \nu_\theta} = {1\over4}\; .  \label{theta} 
\ee
For the purpose of computing $D$ and $\delta$ from Monte Carlo data,
it is much easier to consider the ``wall-wall"
distribution function
\be
P_{w,N}(x) = \sum_{y} P_N(x,y),
\ee
which represents the probability that the endpoint of the walk lies on
a plane at a distance $x$ from the
origin of the walk. In the large-$N$ limit, $P_{w,N}(x)$ has the scaling 
form
\be
P_{w,N}(x) = {1\over \xi} f_w(\rho) \left[1 + O(N^{-\Delta})\right] \qquad 
\rho=\frac{\vert x \vert}{\xi} .
\label{eq-wwEEDF}
\ee
For large $\rho$ we have
\be
f_w(\rho) \approx f_{w,\infty} \rho^{\sigma_w}
     \exp(- D \rho^\delta)\; ,
\label{fwrho-large}
\ee
where $\delta$ is given by \reff{delta}, $D$ is the same constant appearing in 
Eq. \reff{flargerho}, and \cite{CCP-00}
\be
 \sigma_w = {\delta}\left(\nu_\theta - \gamma_\theta + {1\over2}\right) = 
     - {1\over 6}.
\ee

\subsection{Monte Carlo study} \label{sec4.3}

\begin{figure}
\begin{center}
\includegraphics{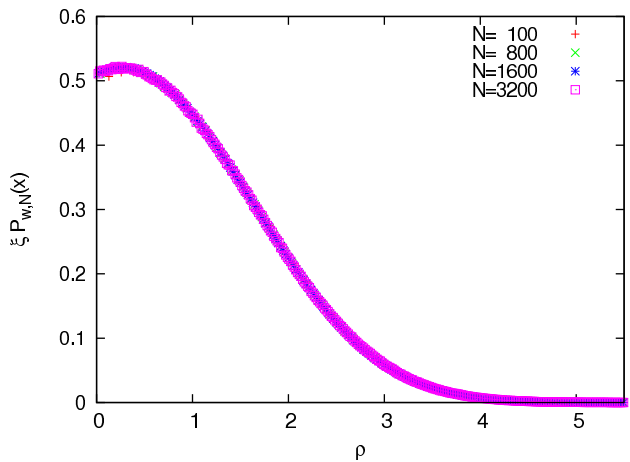}
\caption{The wall-wall EEDF: rescaled combination 
$\xi P_{w,N}(x)$ vs.~$\rho$ for different values of $N$.  }
\label{figure-wallEEDF}
\end{center}
\end{figure}

We studied the EEDF following closely the strategy employed in 
Ref.~\cite{CCP-00} to analyze the same quantity for 
three-dimensional non-interacting SAWs.
Since our runs were performed at $\beta = 0.665 < \beta_\theta$, in principle
we should reweight the MC data to obtain the EEDF at the $\theta$-point.
This correction is apparently negligible compared to the statistical errors, so that 
we directly analyze  
the results for $\beta = 0.665$ without additional corrections.

First, we consider the wall-wall distribution $P_{w,N}(x)$. 
In Fig.~\ref{figure-wallEEDF} we report 
the scaling combination $\xi P_{w,N}(x)$ versus the scaling variable $\rho$.
The scaling is essentially perfect on the scale of the figure, confirming the 
correctness of Eq.~(\ref{eq-wwEEDF}). 
Then, we study the large-$\rho$ behavior with the purpose of 
verifying the asymptotic behavior (\ref{fwrho-large}) and of
determining the constants $D$ and $f_{w,\infty}$. Our data are not 
precise enough to determine $\sigma_w$ and thus we always fix
its value in the numerical analysis.
We fit the data to
\begin{eqnarray}
\ln \xi P_{w,N} &=& \ln f_{w,\infty} - D \rho^\delta, 
\label{fit-fw-large1}\\
\ln (\rho^{1/6} \xi P_{w,N}) &=& \ln f_{w,\infty} - D \rho^\delta.
\label{fit-fw-large2}
\end{eqnarray}
The two fits give correct estimates of $D$ and $\delta$,
but only the second one  provides a correct estimate of $f_{w,\infty}$.
We consider only the data belonging to the range 
$\rho_{\rm min} \le \rho \le \rho_{\rm max}$. 
An upper cut-off $\rho_{\rm max}$ is needed since
scaling corrections and numerical errors 
increase as $\rho$ increases. First, since 
$P_{w,N}(x) = 0$ for $|x| > N$, the deviations from 
(\ref{fit-fw-large1}) and (\ref{fit-fw-large2}) at fixed $N$ 
become infinitely large as $\rho\to N/\xi \sim N^{1-\nu_\theta}$.
Second, since the EEDF decreases rapidly with $\rho$, for $\rho$ large
there is very limited statistics, so that $P_{w,N}(x)$ has a very large error.
But large-$\rho$ data dominate in the fits, providing completely
unreliable estimates of the fit parameters.

\begin{table}
\begin{center}
\begin{tabular}{|ccccc|}
\hline\hline
$N$ & $\rho_{\rm max}$ & $\rho_{\rm min}$ & $\delta$ & $D$ \\ 
\hline
100 & 4 & 2 & 2.517(3)  & 0.127(1) \\ 
    &   & 3 & 2.501(21) & 0.131(5) \\
    & 5 & 2 & 2.466(6)  & 0.142(1) \\ 
    &   & 3 & 2.504(13) & 0.133(3)  \\
    & 6 & 2 & 2.565(14) & 0.117(3) \\ 
    &   & 3 & 2.601(29) & 0.109(6) \\ 
\hline
800 & 4 & 2 & 2.394(12) & 0.148(3) \\ 
    &   & 3 & 2.410(82) & 0.144(20)  \\
    & 5 & 2 & 2.318(9)  & 0.173(3) \\ 
    &   & 3 & 2.305(29) & 0.177(9) \\
    & 6 & 2 & 2.378(20) & 0.153(6) \\ 
    &   & 3 & 2.361(43) & 0.158(13) \\ 
\hline
1600 &4 & 2 & 2.426(5)  & 0.141(1) \\ 
    &   & 3 & 2.390(30) & 0.150(8)  \\
    & 5 & 2 & 2.353(7)  & 0.164(2) \\ 
    &   & 3 & 2.358(21) & 0.162(6) \\
    & 6 & 2 & 2.414(15) & 0.144(4) \\ 
    &   & 3 & 2.430(32) & 0.139(9) \\ 
\hline
3200& 4 & 2 & 2.455(11) & 0.135(2) \\ 
    &   & 3 & 2.423(77) & 0.142(18)  \\
    & 5 & 2 & 2.371(15) & 0.159(4) \\ 
    &   & 3 & 2.375(45) & 0.158(13) \\
    & 6 & 2 & 2.469(31) & 0.132(8) \\ 
    &   & 3 & 2.493(67) & 0.125(16) \\ 
\hline\hline
\end{tabular}
\caption{Estimates of $D$ and $\delta$, obtained by fitting the 
wall-wall EEDF to Eq.~(\ref{fit-fw-large1}).}
\label{table-wallEEDF-1}
\end{center}
\end{table}

Results obtained from fits to Ansatz (\ref{fit-fw-large1})
are reported in Table~\ref{table-wallEEDF-1}, as a function of 
$N = 100, 800, 1600, 3200$. The results for $\delta$ do not 
show systematic dependences  on the fit parameters $\rho_{\rm min}$ and 
$\rho_{\rm max}$ (at least in the range we consider),
while they show a tiny dependence on $N$: apparently, for $N\ge 800$,
$\delta$ increases with increasing $N$. The reason is not fully clear 
but it may be again an effect of the crossover towards the
good-solvent value $\delta = 1/(1 - \nu) = 4$. In any case,
the results are reasonably consistent 
with the theoretical prediction 
$\delta = 7/3 \approx 2.333$. The constant $D$ varies roughly between
0.13 and 0.17 for $N\ge 800$, so that we can estimate $D = 0.15(2)$. 

\begin{table}
\begin{center}
\begin{tabular}{|ccccc|}
\hline\hline
$N$ & $\rho_{\rm min}$ & $\delta$ & $D$ & $\log f_{w,\infty}$ \\
\hline
100 & 2 & 2.527(4) & 0.125(1) & $-$0.640(5) \\ 
    & 3 & 2.545(13) & 0.121(3) & $-$0.670(2) \\ 
\hline
800 & 2 & 2.377(10) & 0.153(3) & $-$0.577(13) \\ 
    & 3 & 2.345(30) & 0.162(9) & $-$0.520(53) \\ 
\hline
1600& 2 & 2.412(7) & 0.145(2) & $-$0.607(9) \\ 
    & 3 & 2.397(22) & 0.149(6) & $-$0.581(37) \\ 
\hline
3200& 2 & 2.431(15) & 0.140(4) & $-$0.620(18) \\ 
    & 3 & 2.417(46) & 0.144(12) & $-$0.598(78) \\ 
\hline\hline
\end{tabular}
\caption{Estimates of $D$ and $\delta$, obtained by fitting the 
wall-wall EEDF to Eq.~(\ref{fit-fw-large2}) with 
$\rho_{\rm max}=5$.}
\label{table-wallEEDF-2}
\end{center}
\end{table}

In order to estimate $f_{w,\infty}$, we cannot neglect the multiplicative 
factor $\rho^{\sigma_w}$ and thus only the results of the second fit 
are relevant. From the data at $N = 1600$ we obtain
\begin{equation} 
\log f_{w,\infty} = -0.60(5), \qquad f_{w,\infty} = 0.55(3),
\end{equation}  
where the error takes into account the estimates obtained
by using all values of $N$.

\begin{figure}
\begin{center}
\includegraphics{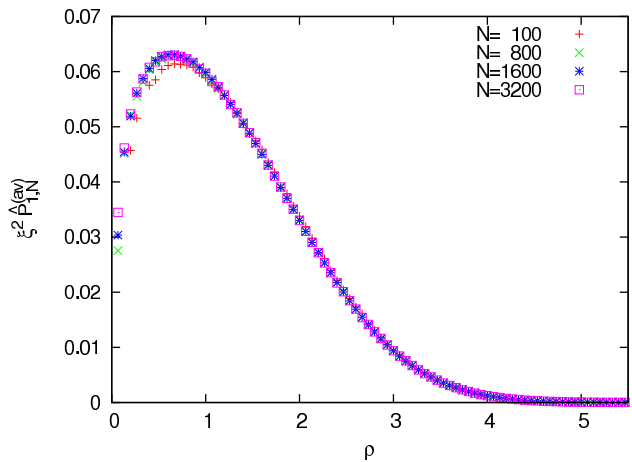}
\caption{Rescaled EEDF $\xi^2 \hat{P}^{\rm (av)}_{1,N}$ with
$\Lambda = 1/15$ vs. $\rho$. }
\label{figure-EEDF}
\end{center}
\end{figure}

Let us consider now the radial distribution $P_N({\bf r})$.
Such a quantity is 
not well suited for a numerical determination of the 
scaling function $f(\rho)$,
because of fluctuations due to the 
lattice structure. In order to average them out, we will employ a
procedure already used in this context in 
Refs.~\cite{Dayantis-Palierne_91,Eizenberg-Klafter_93,CCP-00}. 

We shall consider two different averages
\begin{eqnarray}
\hat{P}^{\rm (av)}_{1,N}(r_{1,n}) &=& 
 {1\over 2 N_{1,n}(r_{1,n})} 
   \sum_{\vec{r}: r^2_{1,n-1} < r^2 \le r^2_{1,n}} {P}_N({\bf r}),
\label{eq-EEDFaverage1}
\\
\hat{P}^{\rm (av)}_{2,N}(r_{2,n}) &=& 
 {1\over 2 N_{2,n}(r_{2,n})} 
   \sum_{\vec{r}: r^2_{2,n-1} < r^2 \le r^2_{2,n}} {P}_N({\bf r}).
\label{eq-EEDFaverage2}
\end{eqnarray}
Here $r_{1,n} = r_0 + n \Delta$ and $r_{2,n}^2 = r_0^2 + n \Delta$, where 
$r_0$ and $\Delta$ are fixed parameters,\footnote{
Procedure (\ref{eq-EEDFaverage1}) corresponds to fixing the
width of the annuli on which the average is computed, while
(\ref{eq-EEDFaverage2}) corresponds to fixing the area (hence the number of points).} 
and $N_{1,n}(r_{1,n})$ and $N_{2,n}(r_{2,n})$ are the number of lattice points
with the same parity\footnote{A point $(x,y)$ is odd (resp. even) if 
$x+y$ is odd (resp. even).} of $N$ that lie in the considered shell.
For practical purposes, we measure $\Delta$ in units of the correlation length:
we define $\Lambda = \Delta/\xi$ and keep it fixed for all values of $N$.
For $\Delta$ fixed, ${P}_N({\bf r})$, $\hat{P}^{\rm (av)}_{1,N}(r)$, and 
$\hat{P}^{\rm (av)}_{2,N}(r)$ have the same scaling behavior as $N\to\infty$.
The same holds for fixed $\Lambda$, as long as $\Lambda \ll 1$.
In Fig.~\ref{figure-EEDF} we report the rescaled EEDF
obtained by using the average \reff{eq-EEDFaverage1}
with $\Lambda=1/15$. All points fall on top of each other, except those 
with $N = 100$ (this is particularly evident for $\rho\lesssim 1.5$). 
This confirms the validity of the scaling relation (\ref{deffrho}).

\begin{table}
\begin{center}
\begin{tabular}{|ccccc|}
\hline \hline
 $N$ & $\rho_{\rm min}$ & $\delta$ & $D$ & $\log f_\infty$ \\ 
\hline
800 & 2 & 2.355(32) & 0.158(10) & $-$2.566(75) \\ 
    & 2.5 & 2.340(42) & 0.163(14) & $-$2.52(11) \\
    & 3 & 2.323(56) & 0.170(20) & $-$2.46(17) \\
    & 3.5 & 2.296(83) & 0.180(32) & $-$2.35(30) \\
\hline
1600& 2 & 2.381(13) & 0.150(4) & $-$2.610(32) \\ 
    & 2.5 & 2.375(17) & 0.153(5) & $-$2.588(47) \\
    & 3 & 2.366(22) & 0.155(7) & $-$2.558(70) \\
    & 3.5 & 2.351(32) & 0.161(11) & $-$2.50(11) \\
\hline
3200& 2 & 2.368(35) & 0.154(11) & $-$2.589(80) \\ 
    & 2.5 & 2.355(47) & 0.159(15) & $-$2.55(12) \\
    & 3 & 2.340(63) & 0.164(21) & $-$2.50(19) \\
    & 3.5 & 2.317(95) & 0.172(34) & $-$2.42(32) \\
\hline\hline
\end{tabular}
\caption{Fit results for the large-distance behavior of the
radial EEDF obtained with the fixed-width 
average (\ref{eq-EEDFaverage1}).
Here $\rho_{\rm max} = 7$, $\Lambda = 1/15$.}
\label{table-EEDF-large-average1}
\end{center}
\end{table}

Let us now again consider the large-$\rho$ behavior.
In order to determine the parameters, we perform fits
of the form 
\be
\log [\xi^2 \hat{P}^{\rm (av)}] = \log f_\infty - D \rho^\delta
\ee
for each $N$ and for several $\rho_{\rm min}$,
$\rho_{\rm max}$, and $\Lambda$. 
Note that in this case
theory predicts $\sigma = 0$ and thus this fit allows us to determine 
$f_\infty$, too.

Results for $N\ge 800$, $\Lambda = 1/15$ and the 
fixed-width average (\ref{eq-EEDFaverage1}) are 
reported in Table \ref{table-EEDF-large-average1}.
The results for $\delta$ are fully consistent with the theoretical
value, while those for $D$ give roughly $D = 0.16(2)$, which is in 
agreement with the estimate obtained by using the wall-wall EEDF.
Estimates using $\Lambda = 1/5$, or obtained by using the average
(\ref{eq-EEDFaverage2}) give similar results. 
As for $f_\infty$ we estimate $\log f_\infty = -2.60(15)$ and
$f_\infty = 0.082(11)$.

More precise estimates of $D$, $f_\infty$, and
$f_{w,\infty}$ are obtained 
by fixing $\delta$ to its theoretical value $\delta = 7/3$.
From the analysis of the wall-wall EEDF,
using the fit function (\ref{fit-fw-large2}), we obtain
\be
D = 0.1668(3) \qquad f_{w,\infty} = 0.625(4),
\ee
while from the radial distribution function we have
\be
D = 0.1656(3) \qquad f_{\infty} = 0.088(2).
\ee
The estimates of $D$ obtained in the two cases differ by two combined 
error bars, indicating that the errors are underestimated by a factor
of at least two. Multiplying all errors by two, 
we end up with the final estimates 
\be
D = 0.1662(6) \qquad f_{w,\infty} = 0.625(8), \qquad f_{\infty} = 0.088(4).
\ee
We finally consider the behavior for $\rho \to 0$, performing fits of the 
form
\be
\log f(\rho) = \log f_0 + \theta \log \rho,
\label{fit-smallrho}
\ee
see Eq. \reff{fsmallrho}.
Since Eq. \reff{fit-smallrho} is valid only for $\rho\to0$
and for $r\to\infty$ (scaling limit) data must be analyzed
in a window $\rho_{\rm min} \le \rho \le \rho_{\rm max}$. 
We find stable results only for $N = 3200$. For lower values of $N$
lattice effects are very strong and Eq.~(\ref{fit-smallrho})
does not describe the low-$r$ data.
If we write $\rho_{\rm min} = n_{\rm min} \Lambda \xi$, we find 
stable results for $n_{\rm min}\gtrsim 1$, 
$0.15\lesssim\rho_{\rm max} \lesssim 0.30$, and $\Lambda$ quite small,
$\Lambda \approx 10^{-2}$. If the parameters are in this range we obtain
\begin{eqnarray}
\theta &=& 0.255(10), \\
f_0 &=&    0.081(2) .
\end{eqnarray}
The result for $\theta$ is in perfect agreement with the
theoretical prediction $\theta=1/4$.
An improved estimate of $f_0$ can be obtained by fixing $\theta$
to its theoretical value. We obtain
\begin{eqnarray}
f_0 = 0.0810(5).
\end{eqnarray}
Finally, we computed the moments
\be
M_{2k,N} = {\sum_{\bf r} r^{2k} P_N({\bf r})\over 
           [\sum_{\bf r} r^{2} P_N({\bf r})]^k },
\ee
see Table \ref{table-moments}.
We extrapolated the results by performing a fit of the form
\be
M_{2k,N} = M_{2k,\infty} + a N^{-\Delta}
\ee
where $M_{2k,\infty}$, $a$, and $\Delta$ are free parameters. 
The results are reported in Table \ref{table-moments}.

\begin{table}
\begin{center}
\begin{tabular}{|cccccc|}
\hline\hline
$N$ & $M_{4,N}$ &  $M_{6,N}$ &  $M_{8,N}$ &  $M_{10,N}$ &  $M_{12,N}$ \\
\hline
100 & 1.778(1) & 4.422(4) & 13.89(2) & 52.2(1) & 226.8(6) \\
800 & 1.815(4) & 4.66(1) & 15.31(7) & 60.8(4) & 281(3) \\
1600 & 1.819(2) & 4.69(1) & 15.49(4) & 61.8(2) & 288(2) \\
3200 & 1.818(4) & 4.68(2) & 15.40(7) & 61.2(4) & 283(3) \\
\hline
$N\to\infty$ & 1.821(4) & 4.70(3) & 15.5(2) & 62(1) & 290(10)\\
\hline\hline
\end{tabular}
\caption{The non-trivial even moments $M_{2k,N}$, $k\le 6$, and the 
corresponding asymptotic values.}
\label{table-moments}
\end{center}
\end{table}

\subsection{Phenomenological expressions}

A phenomenological representation for the function $f(\rho)$
has been proposed by McKenzie and Moore \cite{MM-71} 
and des Cloizeaux \cite{dCJ-book}:
\be
f(\rho) \approx {f}_{\rm ph}(\rho) =
   f_{\rm ph} \rho^{\theta_{\rm ph}}
    \exp\left(-{D}_{\rm ph} \rho^{\delta_{\rm ph}} \right).
\label{eq-frhoapprox}
\ee
Here $\delta_{\rm ph}$ and $\theta_{\rm ph}$ are free parameters,
while ${f}_{\rm ph}$ and ${D}_{\rm ph}$ 
are fixed by the normalization conditions
\reff{norm1} and \reff{norm2}:
\bea
D_{\rm ph} &=& \left\{ {\Gamma[(4 +\theta_{\rm ph})/\delta_{\rm ph}] \over 
        4\,
       \Gamma[(2 +\theta_{\rm ph})/\delta_{\rm ph}] } 
        \right\}^{\delta_{\rm ph}/2}
    \; , \nonumber \\
f_{\rm ph} &=& {\delta_{\rm ph} 
    {D}_{\rm ph}^{(2 +\theta_{\rm ph})/\delta_{\rm ph}} \over 2\pi \,
                   \Gamma[(2 +\theta_{\rm ph})/\delta_{\rm ph}]} \; .
\label{eq-fph}
\eea
In three dimensions in the good-solvent regime this expression describes 
the EEDF quite accurately, even taking $\delta$ and $\rho$ equal to their
theoretical value \cite{CCP-00}.

\begin{figure}
\begin{center}
\includegraphics{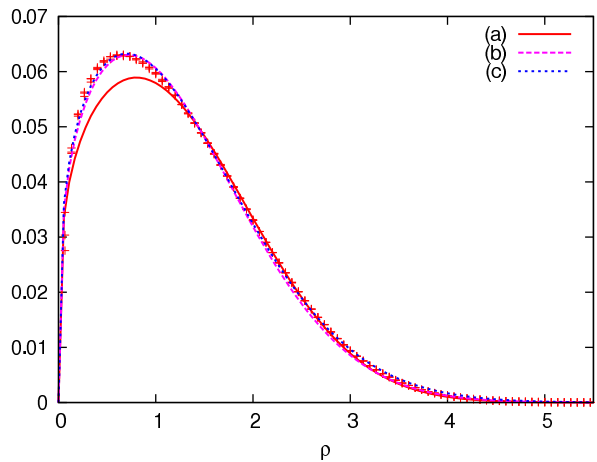}
\caption{The EEDF against several phenomenological approximations:
(a) we set $\delta_{\rm ph}=7/3$ and $\theta_{\rm ph}=1/4$ and use 
Eq.~(\ref{eq-fph}) to fix the constants;
(b) $\delta_{\rm ph}$ and $\theta_{\rm ph}$ are determined by fitting
the data, while the constants are fixed by Eq.~(\ref{eq-fph}) ;
(c) $\delta_{\rm ph}$, $\theta_{\rm ph}$, $D_{\rm ph}$, and 
$f_{\rm ph}$ are obtained by fitting the data.
}
\label{figure-phenomenological}
\end{center}
\end{figure}

In our case, if we use $\theta_{\rm ph} = \theta = 1/4$ 
and $\delta_{\rm ph} = \delta = 7/3$ we obtain for the two constants
\be
D_{\rm ph} = 0.1794, \qquad\qquad f_{\rm ph} = 0.06931,
\ee
which are quite close to the exact results. The resulting curve, curve (a) 
in Fig.~\ref{figure-phenomenological}, reasonably describes the EEDF 
in the large- and small-distance region, but underestimates it in the 
internediate region $0.2 \lesssim \rho \lesssim 1.4$.
As an additional check we can compute the invariant ratios $M_{2k}$.
Using the phenomenological expression 
we obtain $M_{2k,\rm ph} = 1.77,4.39,13.9,53.1, 237$ for 
$k = 2,3,4,5,6$. They are not very much different from the exact 
results reported in Table~\ref{table-moments}, the differences varying 
between 3\% for $k = 2$ and 18\% for $k = 6$. Note that discrepancies increase
as $k$ increases. This is due to the fact that these ratios are increasingly 
sensitive to the large-$\rho$ behavior, and the phenomenological expression
underestimates the EEDF for large $\rho$ since $D_{\rm ph} > D = 0.1662(6)$.

In order to obtain a better approximation, we take $\theta_{\rm ph}$
and $\delta_{\rm ph}$ as free parameters, fixing always 
$D_{\rm ph}$ and $f_{\rm ph}$ by using the normalization conditions
(\ref{eq-fph}). We obtain 
\be
\theta_{\rm ph} \approx 0.282,  \qquad
\delta_{\rm ph} \approx 2.04.
\label{graph_b}
\ee
Correspondingly $D_{\rm ph} = 0.270$, $f_{\rm ph} = 0.0795$.
The resulting phenomenological expression describes better the 
EEDF in the relevant region $\rho \lesssim 5$, 
see Fig.~\ref{figure-phenomenological}, but clearly overestimates
the EEDF in the large-$\rho$ region, given that $\delta_{\rm ph}$
is smaller than $\delta = 7/3 = 2.333$. As a check we again computed
the ratios $M_{2k}$. For $k = 2$ we obtain $M_{4,{\rm ph}} = 1.825$ which agrees
with the correct value $M_4 = 1.821(4)$ and confirms the 
validity of the approximation for $\rho$ not too large.  
However, for $k \ge 3$
the obtained estimates $M_{2k,{\rm ph}}$  are larger than the those 
reported in Table~\ref{table-moments}. For instance we obtain 
$M_{6,{\rm ph}} = 4.96$ and $M_{8,{\rm ph}} = 17.65$, which 
overestimate the correct results by 5\% and 14\%, respectively.

We obtain a slightly better approximation 
if we keep all constants as free 
parameters, relaxing the normalization conditions (\ref{eq-fph}). 
We obtain 
\be
\theta_{\rm ph} = 0.277, \quad \delta_{\rm ph} = 1.95,
\quad D_{\rm ph} = 0.285, \quad f_{\rm ph} = 0.0805.
\label{graph_c}
\ee
The corresponding curve is reported in Fig.~\ref{figure-phenomenological}, as 
graph (c). It cannot be distinguished on the scale of the figure
from graph (b), obtained by using parameters (\ref{graph_b}). This 
is not unexpected, since the parameters are quite close to each other.
For the choice (\ref{graph_c}) of the constants we have
\bea
   \int_0^\infty 2 \pi \rho d\rho\, f(\rho) &=& 1.042\; , \\
   \int_0^\infty 2 \pi \rho^{3} d\rho\, f(\rho) &=& 4.469\; .
\eea
The violations of the normalization conditions are therefore reasonably
small (4\% and 10\% in the two cases).

\subsection{Internal-point distribution function}

As a byproduct of our simulations, we also determined an exponent which is 
related to the internal-point distribution function.
We consider the probability
$P_{N,M}({\bf r})$ that $\omega_M - \omega_0 = {\bf r}$, where 
$\omega_M$ is an internal point, i.e. $M < N$. 
In the limit $N,M\to\infty$, $r\to \infty$ with 
$r N^{-\nu}$ and $M/N$ fixed, we obtain the scaling expression
\be
P_{N,M}({\bf r}) \approx {1\over \xi^2} f_{\rm int}(r/\xi,M/N),
\label{scalingPNM}
\ee
where $\xi^2 = \<R^2_e\>_N/4$ as before. The function $f_{\rm int}(\rho,M/N)$
is nonanalytic for $\rho \to 0$:
\be
f_{\rm int}(\rho,M/N) \sim \rho^{\theta_{\rm int}},
\ee
where the exponent $\theta_{\rm int}$ is independent of $M/N$. 
In two dimensions $\theta_{\rm int}$  has been computed exactly, obtaining 
$\theta_{\rm int} = 5/6$ for noninteracting SAWs and 
$\theta_{\rm int} = 5/12$ at the $\theta$ point
\cite{DS-87}. 

The exponent $\theta_{\rm int}$ 
can be determined by measuring the probability $P_N^{\rm ENN}$ that the 
endpoint is a nearest neighbor of the walk.
Keeping into account that there are $N$ internal points we obtain 
\be
P_N^{\rm ENN} \sim {N\over \xi^2} \xi^{-\theta_{\rm int}} \sim
    N^{-2\nu - \nu\theta_{\rm int} + 1}.
\ee
It should be noted that this expression only takes into account 
``distant" contacts, since the scaling form (\ref{scalingPNM})
is valid only in the limit $r\to \infty$. To this nonanalytic term
we should therefore add the contribution of ``local" contacts, which is 
expected to be an analytic function of $N$. Thus, we obtain the prediction
\be
P_N^{\rm ENN} \approx 
  a + {b\over N} + {c\over N^{\nu(2  + \theta_{\rm int}) - 1}} + \ldots
\ee
At the $\theta$ point this gives
\be
P_N^{\rm ENN} \approx
  a + {b\over N} + {c\over N^{8/21}},
\ee
while for noninteracting SAWs we have 
\be
P_N^{\rm ENN} \approx
  a + {b\over N} + {c\over N^{9/8}}.
\ee
\begin{table}
\begin{center}
\begin{tabular}{|cc|}
\hline\hline
$N$ & $P_N^{\rm ENN}$ \\
\hline
100 & 0.63307(4) \\
800 & 0.71579(10) \\
1600 & 0.73159(6) \\
3200 & 0.74407(10) \\
\hline\hline
\end{tabular}
\caption{Probability that the endpoint is a nearest neighbor of the walk.}
\label{table-Penn}
\end{center}
\end{table}
We have computed $P_N^{\rm ENN}$ at the $\theta$ point\footnote{We 
have also studied $P_N^{\rm ENN}$ for noninteracting SAWs in two 
and three dimensions. In both cases the data are well fitted by 
$a + b/N$, which allows us to conclude that 
$\theta_{\rm int} > 2(1-\nu)/\nu$. In two dimensions this is consistent
with the theoretical prediction, while in three dimensions
it implies $\theta_{\rm int} \gtrsim 1.40$. }
and fitted the results with $a + b/N^{\Delta}$ 
(see Table~\ref{table-Penn}). 
The estimates of $\Delta$ allow us to obtain an estimate of 
$\theta_{\rm int}$:
\be
\theta_{\rm int} = 0.407(11).
\ee
This result 
is in good agreement with the theoretical value $5/12 = 0.4166\ldots$
Moreover, we obtain 
\be
P_\infty^{\rm ENN} = 0.7854(14).
\ee

\section{Conclusions} \label{sec6}

In this paper we present a detailed study of some geometrical properties
of two-dimensional interacting SAWs at the 
$\theta$ point. For this purpose we have generated walks of length up to 
$N = 3200$ at $\beta = 0.665$, which is close to the 
$\theta$ point value $\beta_\theta = 0.6673(5)$.

The main results of this investigation are the following:
\begin{itemize}
\item[(i)] We compute the critical exponents $\nu_\theta$ and $\phi$. 
Our estimate of $\nu_\theta$, $\nu_\theta = 0.570(2)$ is in perfect agreement
with the Coulomb-gas prediction \cite{DS-87} $\nu_\theta = 4/7 \approx 0.571$.
For the exponent $\phi$, we find $\phi = 0.479(6)$ from the analysis of the 
radii, $\phi = 0.436(5)$ from the analysis of their temperature 
dependence, and $\phi = 0. 450(4)$ from the specific heat (errors are 
purely statistical). The somawhat large differences among these estimates 
indicate that the neglected scaling corrections are important. A 
reasonable final estimate would be $\phi = 0.46(3)$, which takes into
account all results with their errors. Thus, we also confirm, although
with limited precision, the theoretical prediction \cite{DS-87}
$\phi = 3/7\approx 0.429$.
\item[(ii)] We compute several invariant ratios involving the radii 
$R^2_g$, $R_m^2$, and $R_e^2$ and, in particular, we verify a 
conjecture of Ref.~\cite{OPBG-94}. For $N\to \infty$, the combination 
$F_N$ defined in Eq.~(\ref{def-F}) vanishes, as it does for noninteracting
SAWs  \cite{CS_1989_JPhysA, CPS_1990_JPhysA}.
\item[(iii)] We discuss the shape of the walks, determining, in particular, the 
average asphericity $\cal A$. We obtain 
\be
   {\cal A} = 0.3726(7),
\ee
where the error is purely statistical. Walks are typically elliptic, the 
ratio of the two axes being 2.11(2). For comparison, note that
for random walks \cite{DE-89} ${\cal A} = 0.3964$, 
while under good-solvent conditions \cite{BCRF-91,vFYMB-09}
${\cal A} = 0.503(1)$.
\item[(iv)] We compute the EEDF. We verify the theoretical predictions
for its small- and large-distance behavior and provide effective approximations
valid in the whole relevant range $r/\xi \lesssim 5$, that is for 
$r/\<R_e\>_N \lesssim 2.5$
(for $r \gtrsim 5\xi$ the EEDF is very small).
\end{itemize}

\section*{Acknowledgements}

M.~Papinutto acknowledges financial support by a Marie Curie European
Reintegration Grant of the 7th European Community Framework
Programme under contract number PERG05-GA-2009-249309.


\end{document}